\documentclass[10pt,journal]{IEEEtran}
\usepackage{graphicx}  
\usepackage{amsmath}
\usepackage{amssymb}

\begin{document}
\title{Electron spin detection in the frequency domain under the interrupted Oscillating Cantilever-driven Adiabatic Reversal (iOSCAR) Protocol}

\author{
	Michael Ting$^*$,
	Alfred~O.~Hero,
	Daniel~Rugar$^\dagger$,
	Chun-yu~Yip,
	and Jeffrey~A.~Fessler,
	\thanks{This work was supported by the DARPA Mosaic program under ARO contract DAAD19-02-C-0055.}
	\thanks{$^\dagger$ D.~Rugar is with the IBM Almaden Research Center, San Jose, CA 95120, USA (rugar@almaden.ibm.com, phone: 408 927 2027, fax: 408 927 2510).}
	\thanks{The other authors are with the Department of Electrical Engineering and Computer Science, University of Michigan, Ann Arbor, MI 48109-2108, USA. $^*$M.~Ting (mting@umich.edu, phone: 734 764 5216), A.~O.~Hero (hero@eecs.umich.edu, phone: 734 763 0564), J.~A.~Fessler (fessler@eecs.umich.edu, phone: 734 763 1434), C.~Y.~Yip (chunyuy@umich.edu, phone: 734 730 2659). Fax: 734 763 8041}
}

%
%


\maketitle

%

\begin{abstract}
Magnetic Resonance Force Microscopy (MRFM) is an emergent technology for measuring spin-induced attonewton forces using a micromachined cantilever. In the interrupted Oscillating Cantilever-driven Adiabatic Reversal (iOSCAR) method, small ensembles of electron spins are manipulated by an external radio frequency (RF) magnetic field to produce small periodic deviations in the resonant frequency of the cantilever. These deviations can be detected by frequency demodulation, followed by conventional amplitude or energy detection. In this paper, we develop optimal detectors for several signal models that have been hypothesized for measurements induced by iOSCAR spin manipulation. We show that two simple variants of the energy detector--the filtered energy detector and a hybrid filtered energy/amplitude/energy detector--are approximately asymptotically optimal for the Discrete-Time (D-T) random telegraph signal model assuming White Gaussian Noise (WGN). For the D-T random walk signal model, the filtered energy detector performs close to the optimal Likelihood Ratio Test (LRT) when the transition probabilities are symmetric. 
\end{abstract}

\section{Introduction}
MRFM is a promising technique for dramatically improving the sensitivity and resolution of magnetic resonance imaging~\cite{sidles91,sidles92,sidles95,rugar01}. One of the immediate goals of this field is to demonstrate the detection of individual electron spins. Successful experiments have already been performed that demonstrate sensitivity on the order of two electron spins for integration times on the order of several seconds~\cite{mamin03}. In order to increase detection speed so that it is suitable for imaging applications, significant advances in force detection, spin manipulation and signal detection are required. In this paper, we address the topic of optimal signal detection.

A general MRFM experiment involves the detection of perturbations of a thin micrometer-scale cantilever whose tip incorporates a submicron ferromagnet. When no electron spins are present, the cantilever acts as a harmonic oscillator. Unpaired electron spins in the sample behave like magnetic dipoles, exerting perturbing forces on the cantilever. Thus, the presence of electron spins can be detected based on measuring the perturbation of the cantilever position from its normal oscillatory behaviour. In particular, the iOSCAR method uses an externally modulated RF field to manipulate the electron spins in such a way as to produce periodic forces on the oscillating cantilever~\cite{mamin03,stipe01}. This results in small changes in the cantilever's natural frequency $\omega_0$. A laser interferometer measures the cantilever displacement; detection of these frequency shifts in the cantilever displacement signal identifies the presence of electron spins.

This methodology can potentially be extended to provide single electron spin sensitivity. Unfortunately, there are a host of practical impediments to achieving this objective. Firstly, the spin-induced changes in $\omega_0$ become extremely small at the single-spin level. Thus, very long integration times are required to detect the single spin signal. However, the integration time is limited by spin relaxation effects that randomly depolarize the electron spin over time. At low temperatures, the relaxation effects are mitigated, which is why current experiments are conducted with temperatures in the millikelvin range. In this low temperature regime, measurements are sensitive to thermal noise from various sources. A major source of thermal noise is the heating of the cantilever by the laser interferometer. Spin detection methodologies must account for spin relaxation effects and low signal-to-noise ratio (SNR).

Four signal models are presented in this paper. The first two are continuous-time (C-T) models while the last two are D-T models. There is a compelling reason for moving to D-T models: they are more tractable to work with in general. The first model is obtained by applying the classical description of an electron spin in a magnetic field~\cite{rugar02}. The result is a set of nonlinear differential equations. The second model is derived using a quasi-static approximation~\cite{berman02}: one obtains a C-T random telegraph process. Neither of the two C-T models have finite-dimensional optimal detector implementations. In~\cite{ting03}, a detector for the first model was proposed that used an extended Kalman Filter (KF)-like state estimator. In~\cite{yipArxiv03,yipAsilomar03}, a hybrid Bayes/Generalized Likelihood Ratio (GLR) detector was developed for the C-T random telegraph model. Both have running times that make a real-time implementation unfeasible at this point. The third model is the generalized D-T equivalent of the C-T random telegraph, and the fourth model is a D-T random walk process. The optimal LRT for these last two models can be derived. Moreover, their running times are $\mathcal{O}(N)$ and $\mathcal{O}(MN)$ respectively, where $N$ is the number of samples per observation, and the number of states in the D-T random walk process is $2M+1$. Surprisingly, it can be shown that there exist simpler detectors, all with $\mathcal{O}(N)$ complexity, that approximate the LRT for the third model, the D-T random telegraph. Simulation shows that one of these simpler forms, the filtered energy detector, has performance that is comparable to the LRT for the D-T random walk under certain conditions.

This paper has two main results. Firstly, the filtered energy detector is approximately asymptotically optimal in the case of the symmetric D-T random telegraph model under the conditions of low SNR, long observation time, and the probability of a transition between consecutive samples $(1-p)$ being small. Secondly, in the general D-T random telegraph model (which includes both symmetric and asymmetric transition probabilities), a hybrid filtered energy/amplitude/energy detector is approximately asymptotically optimal under the conditions of low SNR and long observation time. The outline of this paper is as follows. In Section~\ref{sec:desc}, we briefly review the iOSCAR experiment. This is followed by a discussion in Section~\ref{sec:sig_models} of several iOSCAR signal models. Section~\ref{sec:detectors} consists of reviewing the existing detectors that are commonly used, namely the amplitude and filtered energy detectors, and comparing them with detection schemes that we have developed. Simulation results are presented in Section~\ref{sec:sim_results}.

\section{Description of the iOSCAR experiment} \label{sec:desc}
\begin{figure}[htb]
\begin{center}
\includegraphics[width=3.25in]{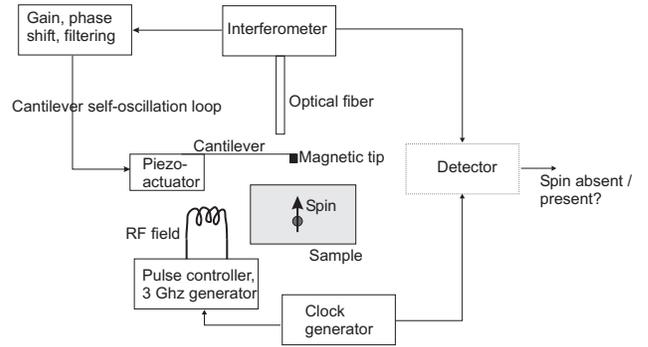} 
\caption{Schematic of the iOSCAR experiment.}
\label{fig:setup}
\end{center}
\end{figure}

A schematic description of the iOSCAR experiment at IBM Almaden is shown in Figure~\ref{fig:setup}. In the current experiment, a submicron ferromagnet is placed on the tip of a cantilever that sits approximately 50 nanometers above a sample. In the presence of an applied RF field, the electron in the sample undergoes magnetic resonance if the RF field frequency matches the Larmor frequency. The latter is proportional to the strength of the cantilever tip's magnetic field. Because the tip field falls off rapidly with distance, only those spins that are within a thin resonant slice will satisfy the condition for magnetic resonance and interact with the cantilever. The resonant slice is located at a certain computable distance away from the cantilever tip. 

If the cantilever is forced into mechanical oscillation by positive feedback, the tip motion will cause the position of the resonant slice to oscillate. As the slice passes back and forth through an electron spin in the sample, the spin direction will be cyclically inverted due to an effect called adiabatic rapid passage~\cite{wago98}. The cyclic inversion is synchronous with the cantilever motion and affects the cantilever dynamics by changing the effective stiffness of the cantilever. Therefore, the spin-cantilever interaction can be detected by measuring small shifts in the period of the cantilever oscillation using a laser interferometer. This methodology has been successfully used to detect small ensembles of electron spins~\cite{mamin03,stipe01}. Signal deconvolution of spin ensemble measurements at different locations above the sample and at different resonant slices can potentially provide single spin resolution~\cite{zuger93}. For more details about iOSCAR, see~\cite{mamin03,stipe01}.

\begin{figure}[!htb]
\begin{center}
\includegraphics[width=3.25in]{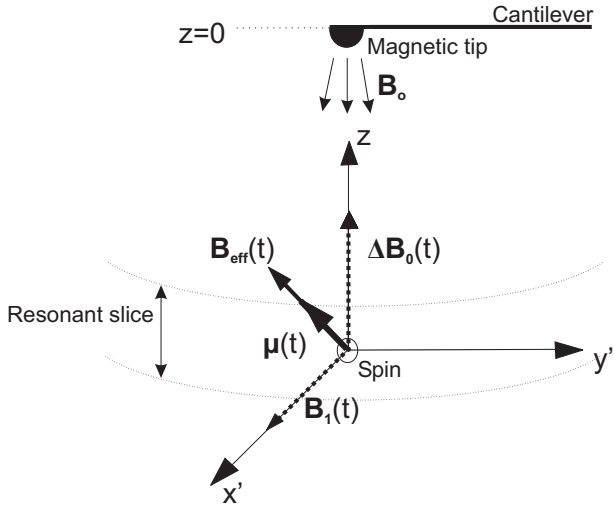} 
\caption{In the coordinate system rotating at $\omega_{\textrm{RF}}$, the off-resonance field $\Delta \vec{B}_{0}(t)$, and therefore the effective field $\vec{B}_{\textrm{eff}}$, oscillate synchronously with the cantilever. Under the spin-lock (anti-spin-lock) assumption, the electron spin aligns with $(-)\vec{B}_{\textrm{eff}}$.}
\label{fig:coord}
\end{center}
\end{figure}

We shall briefly review the setup of the single spin-cantilever interaction framework proposed by Rugar et al.~\cite{rugar02} and Berman et al.~\cite{berman02}. Consider an electron spin in a rotating frame that rotates at the frequency of the applied RF magnetic field $\vec{B}_1(t)$ (see Figure~\ref{fig:coord}). The effective magnetic field $\vec{B}_{\textrm{eff}}(t)$ in this frame is given by
\begin{equation}
\vec{B}_{\textrm{eff}}(t) = B_1(t) \hat{i} + \Delta B_0(t) \hat{k},
\label{eqn:Beff}
\end{equation}
where $\hat{i}$ and $\hat{k}$ are the unit vectors in the $x^\prime$ and $z$ directions of the rotating frame, $B_1(t)$ is the amplitude of the RF magnetic field, $B_0(t)$ is the amplitude of the tip magnetic field at the spin location, and $\Delta B_0(t) = B_0(t) - \omega_{\textrm{RF}}/\gamma$ is the off-resonance field amplitude. The constant $\gamma= 5.6\pi \times 10^{10} \, \textrm{s}^{-1}\textrm{T}^{-1}$ is the gyromagnetic ratio. The spins for which $\omega_{\textrm{RF}}$ approximately equals the Larmor frequency $\omega_L = \gamma B_0(t)$ are said to be in magnetic resonance. Note that $B_0(t)$ is really also a function of space; in our description above, we have fixed the location of the electron so that $B_0(t)$ is just a function of time. Only electrons in a certain slice of the sample will satisfy the magnetic resonance condition, and the position of this slice is a function of the cantilever position. In the rotating frame, $\vec{B}_1(t)$ is a constant vector (except during the skip times which are dictated by the iOSCAR protocol; this will be explained in the next section), and $\Delta B_0(t)$ oscillates synchronously with the cantilever. \textit{If} $\Delta B_0(t)$ varies sufficiently slowly such that the adiabatic criterion
\begin{equation}
\frac{d \Delta B_0(t)}{dt} \ll \gamma B_1^2(t)
\label{eqn:adiabatic}
\end{equation}
is satisfied, the spin can be assumed to remain aligned with either $\vec{B}_{\textrm{eff}}(t)$ or $-\vec{B}_{\textrm{eff}}(t)$. These are the \textit{spin-lock} and \textit{anti-spin-lock} conditions, respectively. 

\section{MRFM signal models} \label{sec:sig_models}
A complete analysis of the spin-cantilever interaction requires a quantum mechanical treatment. Such an analysis is still ongoing. Here, we shall focus on signal models that can be derived from a classical physics framework. A potential weakness of the classical approaches is that they might not adequately characterize the behaviour of the electron, which is subject to quantum effects. However, recent iOSCAR experiments have demonstrated that the key aspects of the classical model are valid. Experimental validation of these models will only be possible once successful detection of a single spin has been demonstrated.
 
The development of the C-T random telegraph model (Model 2) below suggests that almost all of the information pertaining to the presence or absence of a spin is contained in the frequency content of the cantilever position signal $z(t)$. In all of the models, we shall assume that the noise sources are WGN. We shall see that in the presence of a spin, $z(t)$, after being frequency demodulated and translated to baseband, consists of an approximately periodic deterministic square wave and a random signal component. In the absence of the latter, optimal detection can be performed using a matched filter detector. When a random signal component is present, the deterministic part can be cancelled out and we are left with the detection of a random signal in Additive White Gaussian Noise (AWGN).

\subsection{Model 1: Classical C-T model} \label{sec:model1}
The equations of the classical dynamics of a MRFM cantilever interacting with a single electron spin moment are described in~\cite{rugar02}. Considering only the fundamental mode and ignoring the positive feedback term, the interaction is described by:
\begin{align}
\dot{\mu}_x &= \gamma \mu_y(Gz + \delta B_0) \nonumber \\
\dot{\mu}_y &= \gamma \mu_z B_1(t) - \gamma \mu_x(Gz + \delta B_0) \nonumber\\
\dot{\mu}_z &= -\gamma \mu_y B_1(t) \nonumber \\
m\ddot{z} + \Gamma\dot{z} + kz &= G\mu_z + F_n(t) \label{eqn:h1_model1}
\end{align}
where $z(t)$ is the position of the cantilever, $z = 0$ is taken to be the equilibrium position, $m$ is the cantilever's effective mass, and $k$ is the cantilever spring constant. An overhead dot is understood to be differentiation with respect to time. The electron spin moment is given by $\vec\mu(t) = [\mu_x(t) \quad \mu_y(t) \quad \mu_z(t)]^\prime$, and it is known that $\mu_0 = |\vec\mu| = 9.28\times10^{-24} \textrm{J/T}$. $B_1(t)$ is the RF signal which is known, and $F_n(t)$ is WGN which arises due to various noise sources in the experiment, e.g. background thermal noise. The above equations omit the effect of the higher-order modes of the cantilever. This effect can be accommodated by adding more second order equations similar to the last equation in (\ref{eqn:h1_model1}), and with $z_i, i = 2, 3, \ldots$ used in the $i$-th additional equation in place of $z$. Each additional 2nd order equation has a different noise term $F_{ni}(t)$, and the $z$ appearing in the first three equations of (\ref{eqn:h1_model1}) will be replaced by $z + z_2 + \ldots + z_n$, where $n$ is the number of cantilever modes considered. Note that $G \ne 0$, so that when a spin is present, $G\mu_z$ affects the dynamics of $z(t)$, and (\ref{eqn:h1_model1}) is a nonlinear system of differential equations. On the other hand, when a spin is not present, the $G\mu_z$ term vanishes, and we are left with the standard equation of motion for a cantilever, which is:
\begin{equation}
m\ddot{z} + \Gamma\dot{z} + kz = F_n(t) \label{eqn:h0}
\end{equation}

The observable output of the system are samples of the cantilever position $z(t)$ corrupted by observation noise, which is assumed to be AWGN. Define $t_i = iT_s$ to be the time instants at which $z(t)$ is sampled, where $T_s$ is the sampling interval. Model the observation noise as $w_i$, where $w_i$ is a sequence of independent and identically distributed (i.i.d.) Gaussian random variables (r.v.s) with 0 mean and variance $\sigma^2$. Denote the observation sample at time $t_i$ by $y_i$. Then $y_i = z(t_i) + w_i$. The detection problem for this signal model is as follows: given the noisy observations $\vec y = [y_0, \ldots, y_{N-1}]^\prime$, classify the system that generated $\vec y$ as either:
\begin{eqnarray*}
H_0: && \!\!\!\! \vec y \ \ \textnormal{generated by the no-spin system} \\
H_1: && \!\!\!\! \vec y \ \ \textnormal{generated by the spin system}
\end{eqnarray*}

In~\cite{ting03}, we proposed a detector that uses the normal KF and an extended KF-like state estimator for spin detection under Model 1. This detector operates directly on the cantilever position signal. Our focus in this paper is on the detectors for the last two D-T models, and so we shall not make further mention of the dual KF detector. The interested reader is referred to~\cite{ting03} for details and results.

\subsection{Model 2: C-T random telegraph}
In~\cite{berman02}, the classical C-T model (Model 1) is used to obtain a simpler set of equations to describe the spin-cantilever interaction assuming that the cyclic adiabatic inversion condition (\ref{eqn:adiabatic}) holds. A perturbation analysis shows that the cantilever position can then be described by:
\begin{equation}
m\ddot{z}(t) + \Gamma\dot{z}(t) + (k + \Delta k)z(t) = F_n(t) 
\label{eqn:h1_model2}
\end{equation} 

Here, $\Delta k = -\mu G^2/|B_1|$. We note that the cantilever's natural mechanical resonance frequency is $\omega_0 = \sqrt{k/m}$. The shift in the spring constant results in a corresponding shift in $\omega_0$ that is approximately given by $-\frac{1}{2}\omega_0 \frac{\mu G^2}{k |B_1|}$. Define $\Delta\omega_0^\star = \frac{1}{2} \omega_0 |\mu G^2/(k B_1)|$.

With the iOSCAR protocol, $B_1(t)$ is turned off after every $N_{skip}$ cycles over a half-cycle duration to induce periodic transitions between the spin-lock and anti-spin-lock states. This results in $\Delta \omega_0$ alternating between the two values $\pm \Delta \omega_0^\star$. By setting $F_n(t) = 0$ and ignoring the amplitude decay of $z$, the solution to (\ref{eqn:h1_model2}) can be approximated as a frequency-modulated signal:
\begin{equation}
z(t) = Z_0 \; \cos \left[ \omega_0 t + \int_0^t s(\xi) d\xi + \theta \right]
\label{eqn:h1_model2_soln}
\end{equation}
where $Z_0$ is the cantilever oscillation magnitude, $\theta$ is a random phase, and $s(\xi)$ is a square wave that is approximately periodic with non-zero amplitude $\Delta \omega_0^\star$ if a spin is present and zero amplitude otherwise. The reason why $s(\xi)$ is not periodic is because the oscillation period is slightly larger when the cantilever's natural frequency is $\omega_0 + \Delta\omega_0^\star$ as opposed to when it is $\omega_0 - \Delta\omega_0^\star$. However, $|\frac{\Delta\omega_0^\star}{\omega_0}|$ is on the order of $10^{-6}$, which makes $s(\xi)$ approximately periodic. Thus, spin coupling (the presence of a spin) can be detected by frequency demodulating $z(t)$ to baseband and correlating the baseband signal with a known square wave signal derived from $B_1(t)$. 

\begin{figure}[!htb]
\begin{center}
\includegraphics[width=3.0in]{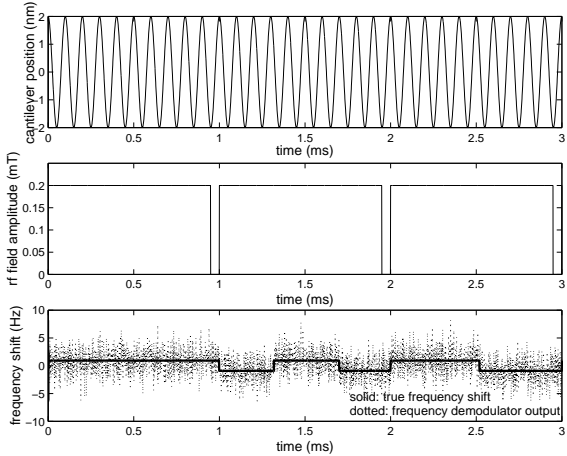}
\caption{Top: Sample of an ideal cantilever position signal. Frequency shifts are not detectable by eye. Middle: Amplitude of sample RF magnetic field, $B_{1}(t)$. It has synchronous half-cycle skips at 1 ms, 2 ms, and 3 ms for the creation of spin state transitions. Bottom: Ideal and noisy outputs of frequency demodulator under the spin presence hypothesis. It has both deterministic transitions due to the RF skips at 1 ms, 2 ms and 3 ms, and random ones due to spin relaxation. The random transitions, $\vec{\tau}$, occur as a Poisson process. The initial polarity is $\phi=1$ for this example.}
\label{fig:signals}
\end{center}
\end{figure}

Unfortunately, the effects of random thermal noise and spin relaxation decorrelate $s(\xi)$ and the square wave signal reference. One model for this decoherence phenomenon is suggested by the Stern-Gerlach experiment~\cite{cohen}: the spins maintain either the spin-lock or anti-spin-lock states, but randomly change polarity during the course of the measurement. This leads to random transitions of $\Delta \omega_0$ between $\pm \Delta \omega_0^\star$, which are assumed to have transition times distributed according to a Poisson process with a rate of $\lambda$ spin reversals/sec. Note that correlating the frequency demodulator output with the known square wave signal, as was described in the previous paragraph, has the effect of cancelling out the deterministic transitions in $\omega_0$. What remains after correlation are the random transitions, and as the transition times are generated by a Poisson process, the resultant signal takes the form of a so-called random telegraph process~\cite{starkNwoods}. See Figure~\ref{fig:signals}.

More specifically, let the baseband output of the frequency demodulator and correlator be denoted by $y(t)$. Let $[0,T]$ be the total measurement time period over which the correlator integrates the measurements, and let $\vec\tau = \{\tau_{i}\}, i=1, \ldots,\mathcal{K}$, be the time instants within this period at which random spin reversals occur. As $\vec{\tau}$ are the arrival times of a Poisson process with intensity $\lambda$, $\mathcal{K}$ is a Poisson random variable with rate $\lambda T$. Thus, the random telegraph model is: $y(t)=s(t)+w(t)$ where $w(t)$ is AWGN with variance $\sigma_{w}^{2}$, and $s(t)$ is a random telegraph signal containing only the random transitions. The detection problem for this model is to design a test between the two hypotheses:
\begin{align}
H_0 \textrm{ (spin absent) } &: \quad y(t) = w(t) \nonumber \\
H_1 \textrm{ (spin present) } &: \quad y(t) = s(t) + w(t)
\label{eq:nine}
\end{align}
for $t\in[0,T]$.  

A hybrid Bayes/GLR detector was previously developed for the C-T Random Telegraph model (Model 2)~\cite{yipArxiv03,yipAsilomar03}. Essentially, the detector is the LRT but with the unknown initial phase $\phi$ averaged out and the Maximum Likelihood (ML) estimate of $\vec\tau$ and $N$ used. The test statistic is  
\begin{eqnarray}
\ln \Lambda(y) & \!\!\!\!=\!\!\!\! & \max_{\vec\tau,N} 
	\bigg\{ \ln \cosh \bigg[
		\frac{1}{\sigma_{w}^{2}}
		\int_{0}^{T} \!y(t)s^{+}(t;\vec\tau,N)dt
	\bigg] \bigg\} \nonumber  \\ & \!\!\!\!-\!\!\!\!
&\frac{1}{2\sigma_{w}^{2}}\int_{0}^{T}\!(s^{+}(t;\vec\tau,N))^{2} dt
\label{eqn:CM}
\end{eqnarray}
where $s^{+}(t;\vec\tau,N)$ is the synthesized telegraph signal having initial polarity $\phi=1$ (since $E_{\phi}[\cdot]$ has been taken) and parametrized by $\vec\tau$ and $N$. As Model 2 is C-T, the parameter space of $\{\vec\tau, N\}$ is infinite-dimensional. In~\cite{yipArxiv03,yipAsilomar03}, a Gibbs sampler was implemented to efficiently search the parameter space.

\subsection{Model 3: Discrete-Time Random Telegraph}
Model 3 is the generalized D-T equivalent of Model 2. Here, we shall likewise treat $\{y_i\}_{i=0}^{N-1}$ as samples of the baseband output of the frequency demodulator and correlator. The D-T random telegraph signal is a D-T Markov chain, and will be denoted by $\zeta_i$, where $\zeta_i \in \{+A,-A\}, 0 \le i \le N-1$, and $\zeta_0$ is equally likely to be either $\pm A$, $A = \Delta \omega_0^\star$. The transition probabilities of $\zeta_i, i \ge 1$ are as follows:
\begin{equation}
P(\zeta_i|\zeta_{i-1}) = \left\{ \begin{array}{ll}
p & \zeta_i = \zeta_{i-1} = A \\
1-p & \zeta_i = -A, \zeta_{i-1} = A \\
q & \zeta_i = \zeta_{i-1} = -A \\
1-q & \zeta_i = A, \zeta_{i-1} = -A
\end{array} \right.
\label{eqn:model3_trans_probs}
\end{equation}

We restrict $0 < p,q < 1$. If $p = q$, we say that the transition probabilities are symmetric, and when $p \neq q$, we shall say that they are asymmetric. For the symmetric case, we can match the C-T model to the D-T model by equating the expected number of transitions of the Poisson process to that of the Markov chain. This results in $p = 1 - T_s \lambda$. Recall that $T_s$ is the sampling time interval and $\lambda$ is the expected number of transitions per second. Define the signal vector $\vec\zeta = [\zeta_0, \ldots, \zeta_{N-1}]^\prime$ and the noise vector $\vec w = [w_0, \ldots, w_{N-1}]^\prime$. We shall model the $w_i$'s as i.i.d. Gaussian r.v.s with mean 0 and variance $\sigma^2$. The detection problem is then to decide between:
\begin{align}
H_0 \textrm{ (spin absent) } &: \quad \vec y = \vec w \nonumber \\
H_1 \textrm{ (spin present) } &: \quad \vec y = \vec\zeta + \vec w 
\label{eqn:model3_hypo}
\end{align}

\subsection{Model 4: Brownian Motion (BM) on a sphere}
The D-T random telegraph signal model (Model 3) characterized the spin decoherence as producing random transitions of $\Delta \omega_0$ between $\pm \Delta \omega_0^\star$. This is a consequence of the assumption that the electron spin maintains either the spin-lock or anti-spin-lock states. An alternative model has been proposed which characterizes the spin as being BM on a sphere in $\mathbf{R}^3$~\cite{fredkin01}. Only the $z$-component of the spin $\vec{\mu}$ is measured by the cantilever: refer to the last equation of (\ref{eqn:h1_model1}). As an approximation, we shall characterize the effect on $\Delta\omega_0$ as producing a one-dimensional random walk confined to the interval $I = [-\Delta\omega_0^\star,\Delta\omega_0^\star]$. Discretize $I$ into $(2M+1)$ states using a step size of $s$, where $M\in\mathbf{Z}$ and $M,s>0$. Define $I_d = [-Ms,Ms]$ and $\zeta_i$ to be the D-T random walk restricted to $I_d$. Henceforth, we shall refer to this model as the D-T random walk model, keeping in mind that it is an approximation derived from assuming that the spin behaves like BM on a sphere. Associate with $\zeta_i$ the probability transition matrix $P$, so that $P_{jk} = P[ \zeta_i=(k-M)s | \zeta_{i-1}=(j-M)s ], 0 \le j,k \le 2M$, for $i \ge 1$. $P$ is defined so that, at each time step, $\zeta_i$ changes by either $\pm s$. We shall assume reflecting boundary conditions in order to keep $\zeta_i$ in $I_d$, and $\zeta_0$ is equally likely to be either $\pm s $.

The detection problem is now to test (\ref{eqn:model3_hypo}) when $\vec\zeta$ is modelled by a random walk. Note that the D-T random walk model can be regarded as a multi-state generalization of the D-T random telegraph model. In the limit as $s \rightarrow 0, M \rightarrow \infty$, the random walk converges to BM over the interval $I$~\cite{starkNwoods}.

\section{Frequency domain detection strategies} \label{sec:detectors}
The detectors considered here can be placed into three categories: D-T versions of existing detectors that are currently in use; D-T LRTs for Models 3 and 4; and approximations to the LRT for Model 3. The LRT is a most powerful (MP) test that satisfies the Neyman-Pearson criterion: it maximizes the probability of detection ($P_D$) subject to a constraint on the probability of false alarm ($P_F$)~\cite{vantreesv1}, which is set by the user. This gives us a benchmark with which to compare the other detectors tests for Models 3 and 4. When the random transition times are known, the optimal LRT is the matched filter, called the omniscient matched filter (MF) in this paper. Although unimplementable in reality, the MF detector provides an absolute upper bound when comparing the various detectors' Receiver Operating Characteristic (ROC) curves.

The framework for the detectors is depicted in Figure~\ref{fig:baseband} below. Note that in Figure~\ref{fig:baseband}, the C-T quantity $y(t)$ is shown as an input to the statistic generator; however, the detectors in this section operate on the sampled values $\{y_i\}$.
\begin{figure}[!htb]
\begin{center}
\includegraphics[width=3in]{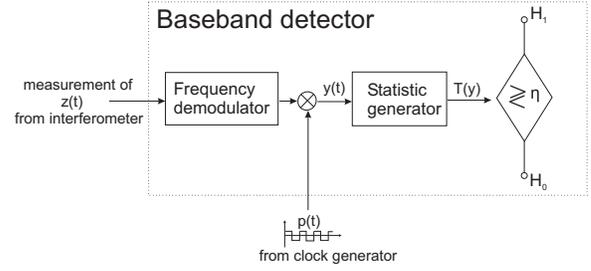}
\caption{Baseband detector frequency demodulates the interferometric signal, correlates the output against a square wave $p(t)$ whose transitions are synchronous with the turn-off times of the RF field $B_1(t)$, and generates a test statistic (e.g., accumulated squared frequency deviations), for detecting the presence of a spin.}
\label{fig:baseband}
\end{center}
\end{figure}

Define the SNR to be $\textrm{SNR} = (\lim_{i\rightarrow\infty} E[\zeta_i^2])/\sigma^2$, where $\zeta_i$ is either the D-T random telegraph or random walk process. I.e.~the SNR is the ratio of the steady-state expected energy of $\zeta_i$ to the noise variance. For the former process, $\textrm{SNR} = A^2/\sigma^2$. Let the SNR in dB be $\textrm{SNR}_{\textrm{dB}} = 10 \, \log_{10} \textrm{SNR}$.

\subsection{Amplitude, energy, filtered energy detectors}
The D-T amplitude detector is
\begin{equation}
\left| \frac{1}{N} \sum_{i=0}^{N-1} y_i \right|
\begin{array}{c} H_1 \\ \gtrless \\ H_0 \end{array} \eta
\label{eqn:det_amplitude}
\end{equation}
where $\eta$ is set to satisfy the constraint on $P_F$. The threshold $\eta$ can be empirically determined by testing (\ref{eqn:det_amplitude}) under $H_0$. It is equivalent to the optimal LRT under the assumption that $y_i$ is the sum of a random constant and i.i.d. WGN. This assumption would be true if there were no random transitions in $\Delta \omega_0$. Under such a situation, the amplitude detector is a MF detector. However, as the number of random transitions in $y_i$ increases, the performance of the amplitude detector degrades. An alternative test statistic is the D-T energy detector, i.e. the sum of the squares of the $\{y_i\}$ instead of the magnitude of the sum in (\ref{eqn:det_amplitude}). As the signal and noise are assumed to be independent, under $H_1$, one would expect $\{y_i\}$ to have a higher energy on average than under $H_0$. This can be reliably detected under a sufficiently high SNR. A natural improvement to the energy detector is possible by pre-filtering $\{y_i\}$ over the signal bandwidth. As the signal $\{\zeta_i\}$ is baseband, a Low Pass Filter (LPF) is appropriate. In particular, we shall use one of the simplest possible LPFs: a first-order, single-pole filter given by
\begin{equation}
H_{\textrm{LP}}(z) = \frac{1-\alpha}{2}\frac{1+z^{-1}}{1-\alpha z^{-1}}
\label{eqn:lpf}
\end{equation}
where we require $|\alpha| < 1$ for stability~\cite{mitra}. The time constant $\alpha$ should be chosen based on the bandwidth of the signal; if $\omega_c$ is the desired -3dB bandwidth of the filter, set $\alpha = (1-\sin\omega_c)/\cos\omega_c$. The -3dB bandwidth depends on the mean number of transitions, i.e. $\lambda$ or $\frac{1-p}{T_s}$ in Models 2 and 3 respectively. In practice, a bank of LPFs with different $\alpha$'s are used to perform detection. Let $a_i=y_i*h_i$. That is, $a_i$ is the sequence obtained by convolving $y_i$ with $h_i$, where $h_i$ is the impulse response of whatever filter we choose to use. The energy and filtered energy detector can then be expressed as
\begin{equation}
\sum_{i=0}^{N-1} a_i^2 
\begin{array}{c} H_1 \\ \gtrless \\ H_0 \end{array} \eta
\label{eqn:det_fe}
\end{equation}
where for the energy detector, $h_i$ is taken to be the unit impulse function $\delta[i]$, while for the filtered energy detector, $h_i = h_{\textrm{LP}}[i]$, the impulse response of $H_{\textrm{LP}}(z)$ in (\ref{eqn:lpf}).

We note that the running time for the amplitude, filtered energy, and energy detectors is $\mathcal{O}(N)$.
 
\subsection{Optimal LRT detectors and their approximations}
One can derive the LRTs for the D-T signal models (Models 3 and 4). Consider first the D-T random telegraph model (Model 3): define $R_k(S) = P( \zeta_k = S| Y_{k-1}, \ldots, Y_0 )$, where $S \in \{\pm A\}$ and $k \ge 1$. Let $\mathcal{N}(x;\mu,\sigma^2) = \frac{1}{\sqrt{2\pi}\sigma} \textrm{exp} \left[ -\frac{(x-\mu)^2}{2\sigma^2} \right]$ and $\gamma_{S_1, S_2} = P( S_1 \rightarrow S_2)$ be the probability that the signal $\zeta_i$ goes from $S_1$ in the current time step to $S_2$ in the next with $S_1, S_2 \in \{\pm A\}$. There exists a recursive formula for $R_k(S)$.
\begin{align}
R_k(S) &= \gamma_{A,S} \underbrace{
		\frac{ e^{\frac{A}{\sigma^2} y_{k-1}} R_{k-1}(A) }{
		e^{ \frac{A}{\sigma^2} y_{k-1}} R_{k-1}(A) +
		e^{ -\frac{A}{\sigma^2} y_{k-1}} R_{k-1}(-A) } }_{\bigstar}
		\nonumber \\
	&+ \gamma_{-A,S} (1 - \bigstar) \label{eqn:R_simpler}
\end{align}
for $k \ge 1$ and with initial conditions $R_0(A) = R_0(-A) = 1/2$. With this, one can derive $f(\vec y;H_1)$, the probability density function (pdf) of $\vec y$ under $H_1$. Let $f(\vec y;H_0)$ be the pdf of $\vec y$ under $H_0$. Define $y^{(k)} = (y_k, \ldots, y_0)$ for $k \ge 0$. Now,
\begin{align}
f(\vec y;H_1) &= f(y_{N-1}|y^{(N-2)};H_1) \cdot \nonumber \\ 
	&f(y_{N-2}|y^{(N-3)};H_1) \cdots f(y_1|y_0;H_1) f(y_0;H_1)
\label{eqn:dtrt_fh1}
\end{align}
and 
\begin{align}
f(y_k|y^{(k-1)};&H_1) = R_k(A)\mathcal{N}(y_k;A,\sigma^2) \nonumber \\
	& + R_k(-A)\mathcal{N}(y_k;-A,\sigma^2), \; k \ge 1
\label{eqn:dtrt_fh1_indv}
\end{align}
	
With (\ref{eqn:dtrt_fh1}) and (\ref{eqn:dtrt_fh1_indv}), the log LRT expression is:
\begin{align}
\ln \, \Lambda(\vec y) &= 
	\ln \, \frac{ f( \vec y;H_1 ) }{ f( \vec y;H_0 ) } \nonumber \\
	&= \sum_{k=0}^{N-1} \ln \left[
	R_k(A) e^{\frac{A}{\sigma^2} y_k} +
	R_k(-A) e^{-\frac{A}{\sigma^2} y_k} \right]
	\begin{array}{c} H_1 \\ \gtrless \\ H_0 \end{array} \eta
\label{eqn:2state_lrt}
\end{align}
We see that, at each time step, the log LRT incorporates information from the present observation in $\textrm{exp}(\pm Ay_k/\sigma^2)$ and information from the past observations in $R_k(\pm A)$. The running time of (\ref{eqn:2state_lrt}) is $\mathcal{O}(N)$, where $N$ is the number of observations.

Under the regime of low SNR and long observation times ($N \gg 1$), the second-order expansion of (\ref{eqn:2state_lrt}) is approximately equal to the hybrid filtered energy/amplitude/energy detector:
\begin{equation}
\sum_k a_k^2 +
	\frac{1-\alpha^2}{2\alpha} C_{\textrm{I}} \sum_k y_k +
	\frac{1-\alpha^2}{2\alpha} C_{\textrm{II}} \sum_k y_k^2
\label{eqn:energy_hybrid}
\end{equation}
where the constants $C_{\textrm{I}}$ and $C_{\textrm{II}}$ are given in the appendix. Here, $a_k = y_k * h_{\textrm{LP}}[k]$, i.e.~the output of the observations convolved with the LPF in (\ref{eqn:lpf}). What this means is that in the aforementioned regime, we expect the hybrid detector to have performance similar to the optimal LRT test. When $p=q$, the second-order expansion of the LRT is approximately equal to the filtered energy detector for values of $p$ close to $1$. See the appendix for more details. In light of the running times for the filtered energy, energy, and amplitude detectors, the complexity of (\ref{eqn:energy_hybrid}) is also $\mathcal{O}(N)$.
 
Next, consider Model 4, the D-T random walk. Define the vectors:
\begin{align*}
\overrightarrow{R}_k = \left[ \begin{array}{c} R_k(-Ms) \\ \vdots \\ 
	R_k(0) \\ \vdots \\ R_k(Ms) \end{array} \right],
\overrightarrow{W}_k = \left[ \begin{array}{c} f_w( y_k + Ms ) \\ \vdots \\
	f_w( y_k ) \\ \vdots \\ f_w( y_k - Ms) \end{array} \right]
\end{align*}
where $f_w(\cdot) = \mathcal{N}(\cdot;0,\sigma^2)$. If $\vec a = (a_1, \ldots, a_n)'$ and $\vec b = (b_1, \ldots, b_n)'$, define the operation $\vec a * \vec b = (a_1 b_1, \ldots, a_n b_n)'$ and $\vec a \cdot \vec b = \sum_i a_i b_i$ (i.e. $\vec a \cdot \vec b$ is the dot product). Let $Q = P^\prime$. As in the LRT for Model 3, there exists a recursive formula for $\overrightarrow{R}_k$:
\begin{equation}
\overrightarrow{R}_k = \frac{ Q( \overrightarrow{W}_{k-1} * 
	\overrightarrow{R}_{k-1} ) }{ \overrightarrow{W}_{k-1} \cdot
	\overrightarrow{R}_{k-1} }
\label{eqn:R_rw}
\end{equation}
for $k \ge 1$ and with the initial condition $\overrightarrow{R}_0 = \frac{1}{2}(\vec{e}_M + \vec{e}_{M+2})$, where $\{\vec{e}_i\}$ are the standard basis vectors for $\mathbf{R}^{2M+1}$. The LRT for the D-T random walk (details are given in the appendix) can be expressed as
\begin{equation}
\Lambda(\vec y) = \prod_{k=0}^{N-1} \frac{
	\overrightarrow{R}_k \cdot \overrightarrow{W}_k }{
		\vec{e}_{M+1} \cdot \overrightarrow{W}_k }
\label{eqn:lrt_rw}
\end{equation}

The running time of the LRT for the D-T random walk is $\mathcal{O}(M^2N)$ for a general matrix $Q$. If $Q$ is tridiagonal, as is the case for the random walk model, the running time is $\mathcal{O}(MN)$.

\section{Simulation results} \label{sec:sim_results}
The objective in this section is to compare all of the detection methods discussed in this paper. The class of LRT detectors is optimal for their respective signal models, and provides a good comparison benchmark. Comparison of the various detectors is done using ROC curves, which is a plot of probability of detection ($P_D$) vs.~probability of false alarm ($P_F$), and power curves, which is a plot of $P_D$ vs.~SNR at a fixed $P_F$. Some of the parameters used in the simulation of Models 3 and 4 are as follows: $k = 10^{-3} \;\textrm{N m}^{-1}$, $\omega_0 = 2\pi \cdot 10^4 \;\textrm{rad s}^{-1}$, $B_1 = 0.2 \;\textrm{mT}$, $G = 2 \cdot 10^6 \;\textrm{T m}^{-1}$. The sampling period was $T_s = 1 \;\textrm{ms}$, and signal durations of $T = 60\;\textrm{s}$ and $T = 150\;\textrm{s}$ were used. The performance of the detectors varies as a function of $T$; in general, a larger $T$ results in better performance. Realistic values of $T$ are several orders of magnitude larger. Nevertheless, the comparative results obtained from using the two values of $T$ above are representative of larger values. Indeed, our approximations to the optimal detectors improve with larger $T$. 

\subsection{D-T random telegraph model (Model 3)}
\begin{figure}[!htb]
\begin{center}
\includegraphics[width=3in]{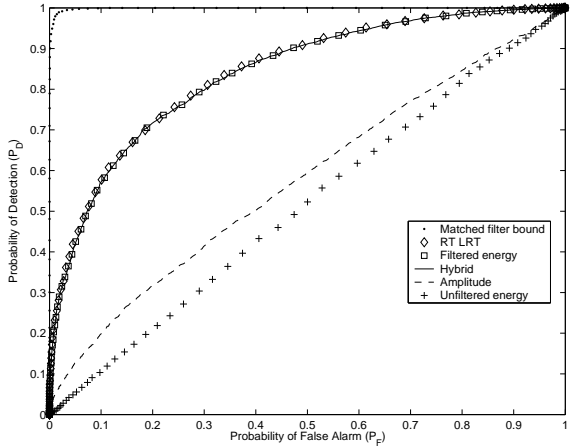}
\caption{Simulated ROC curves for the D-T random telegraph model (Model 3) with symmetric transition probabilities at SNR = \mbox{-35 dB}, $T$ = 60 s, and $\lambda = 0.5 \;\textrm{s}^{-1}$ for the omniscient matched filter, D-T random telegraph LRT, filtered energy, hybrid, amplitude, and unfiltered energy detectors. The RT-LRT is the optimal detector for this model.}
\label{fig:roc_mod3_1}
\end{center}
\end{figure}

\begin{figure}[!htb]
\begin{center}
\includegraphics[width=3in]{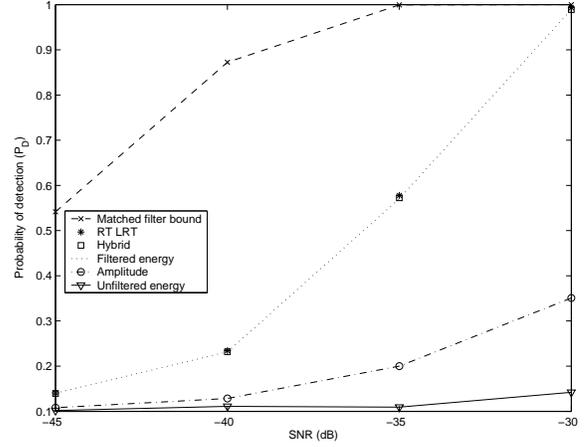}
\caption{Simulated power curves ($P_D$ vs. SNR) for the D-T random telegraph model (Model 3) with $P_F$ fixed at 0.1 and $\lambda = 0.5 \;\textrm{s}^{-1}$, $T$ = 60 s. The RT-LRT is the optimal detector for this model.}
\label{fig:power_mod3}
\end{center}
\end{figure}

First, consider Model 3, the D-T random telegraph. Figure~\ref{fig:roc_mod3_1} depicts the simulated ROC curves at SNR = \mbox{-35 dB}, $\lambda = 0.5 \;\textrm{s}^{-1}$, and with symmetric transition probabilities ($p = q$). With $T_s=1\;\textrm{ms}$, this results in $p=q=0.9995$. We examine the matched filter, D-T random telegraph LRT (RT-LRT), filtered energy, hybrid, amplitude, and unfiltered energy detectors. The RT-LRT, filtered energy, and hybrid detector curves are virtually identical, which confirms our previous analysis. We note that the unfiltered energy and amplitude detectors have performance that is poorer than the RT-LRT, as it should be since this is the optimal detector. The unfiltered energy detector has the worst performance out of the five detector methods considered, and we shall see that this is almost always the case. Lastly, the omniscient MF detector has the best performance. Again, that is consistent with our expectations. We generated a power curve over a range of SNR under the same conditions as in Figure~\ref{fig:power_mod3} with a fixed $P_F = 0.1$ The RT-LRT, filtered energy, and hybrid detector have similar performance from \mbox{-30 dB} to \mbox{-45 dB}. With this particular value of $P_F$ and $\lambda$, the RT-LRT, filtered energy, and hybrid detector perform from \mbox{5 dB} to \mbox{10 dB} worse than the MF detector. Although the amplitude detector has worse performance than the RT-LRT and filtered energy detector, all three have comparable performance at \mbox{-45 dB}. 

\begin{figure}[!htb]
\begin{center}
\includegraphics[width=3in]{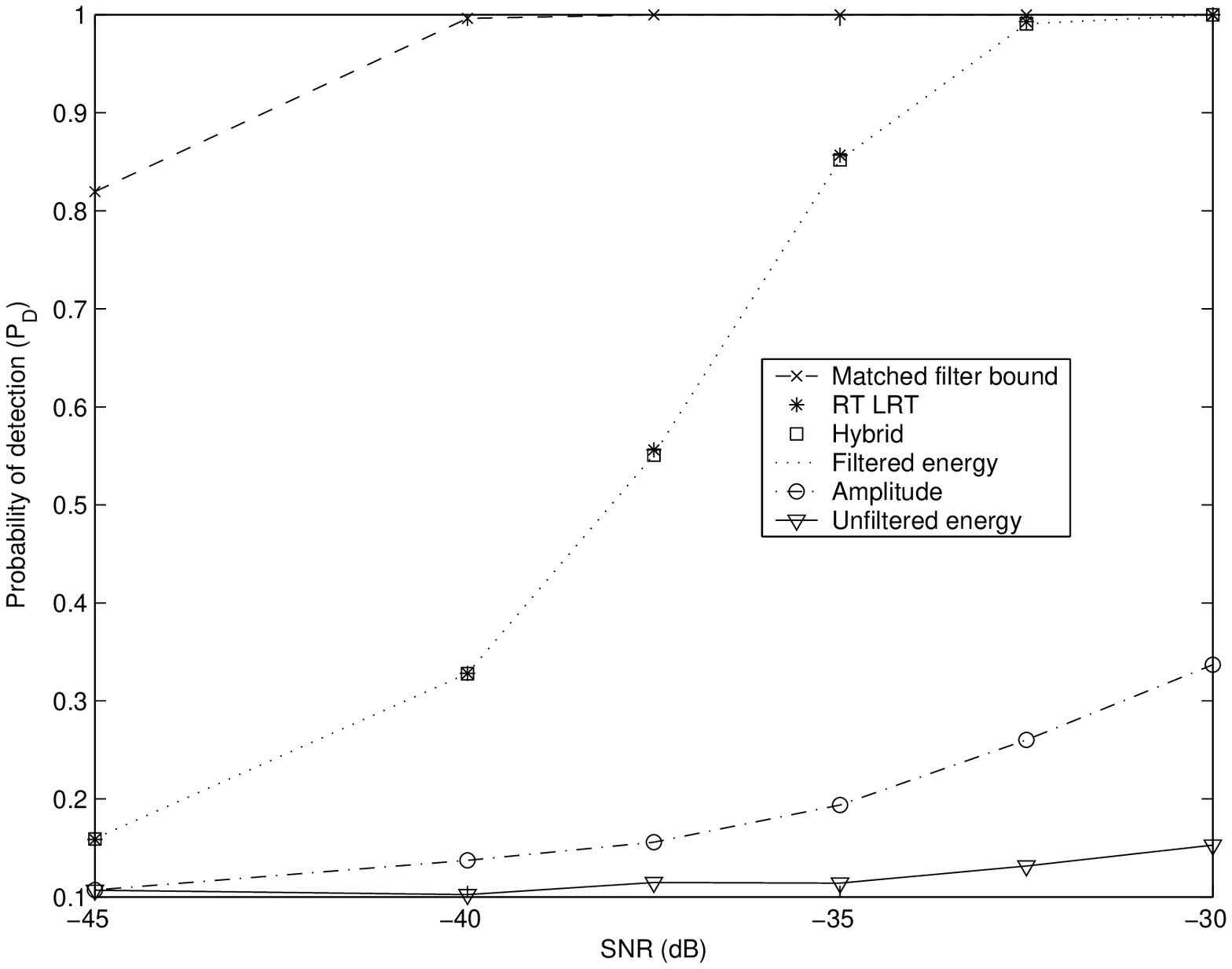}
\caption{Simulated power curves ($P_D$ vs. SNR) for the D-T random telegraph model (Model 3) with $P_F$ fixed at 0.1 and $\lambda = 0.5 \;\textrm{s}^{-1}$, $T$ = 150 s. The RT-LRT is the optimal detector for this model.}
\label{fig:power_mod3_longer}
\end{center}
\end{figure}

Figure~\ref{fig:power_mod3_longer} shows the power curve generated using the bigger value of $T$ = 150 s. Again, the RT-LRT, filtered energy, and hybrid detectors have the same performance from \mbox{-30 dB} to \mbox{-45 dB}. Note that the values of $P_D$ have increased as compared to Figure~\ref{fig:power_mod3}.

\begin{figure}[!htb]
\begin{center}
\includegraphics[width=3in]{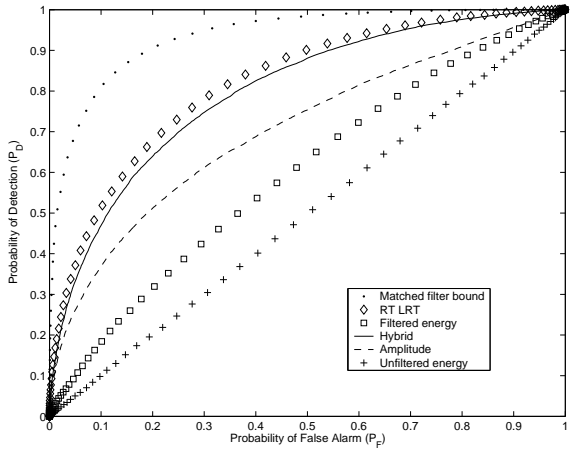}
\caption{Simulated ROC curves for the D-T random telegraph model (Model 3) with asymmetric transition probabilities ($p=0.9998,q=0.9992$) at SNR = \mbox{-45 dB}, $T$ = 150 s. }
\label{fig:roc_mod3_asym}
\end{center}
\end{figure}

In the interest of space, ROC curves for a different value of $\lambda$ will not be shown. However, performance degrades as $\lambda$ increases. In any case, the curves for the RT-LRT and filtered energy detector are similar. Before moving on, we would like to present an asymmetric case where $p \neq q$: set $p=0.9998,q=0.9992$. The ROC curves are presented in Figure~\ref{fig:roc_mod3_asym}. There is a noticeable difference between the curves of the RT-LRT and filtered energy detectors. The hybrid detector's curve is slightly below that of the LRT, and it is better than that of the filtered energy detector. In fact, the filtered energy detector has worse performance than the amplitude detector.  An asymmetry in $p, q$ leads to a non-zero mean signal, which might be why the amplitude detector's performance improves. Indeed, for the D-T random telegraph model, $\lim_{i \rightarrow \infty} E[\zeta_i] = A\frac{p-q}{2-p-q} = 0.6A$ for the values of $p$ and $q$ used here. There would therefore be larger segments of the signal that look constant. Asymmetric transition probabilities can arise in some situations, e.g. experimental conditions or the feedback cooling of spins protocol proposed by Budakian~\cite{rugar_oct03_mosaic_review}.

\begin{figure}[!htb]
\begin{center}
\includegraphics[width=3in]{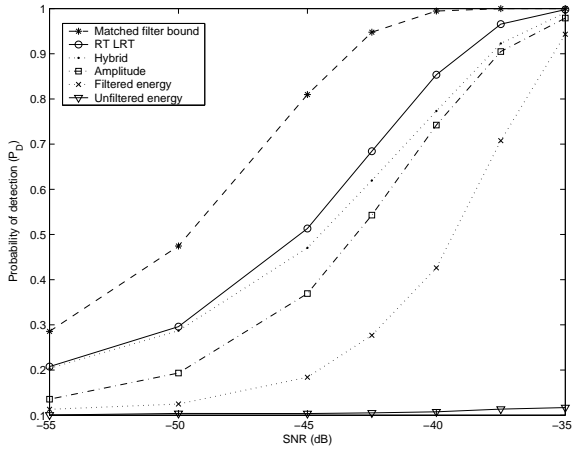}
\caption{Simulated power curves ($P_D$ vs. SNR) for the D-T random telegraph model (Model 3) with $P_F$ fixed at 0.1, $p=0.9998,q=0.9992$, and $T$ = 150 s. The RT-LRT is the optimal detector for this model.}
\label{fig:power_mod3_asym}
\end{center}
\end{figure}

We generated a power curve from SNR = \mbox{-55 dB} to \mbox{-35 dB} for the asymmetric case in Figure~\ref{fig:power_mod3_asym}. It seems that a larger value of $T$ is required when $p \neq q$ for the hybrid filtered energy/amplitude/energy detector to approximate the optimal LRT, hence why we used $T$ = 150 s for simulations of the asymmetric random telegraph model. The hybrid detector has better performance than the amplitude and filtered energy detectors. It has performance that is comparable to the RT-LRT for lower SNR values.

\subsection{D-T random walk model (Model 4)}
For the D-T random walk model, the probability transition matrix $P$ is tridiagonal. Suppose for the moment that $M$ is even. Recall that the random walk $\zeta_i$ is confined to the interval $[-Ms,Ms]$. Define the lower-quartile transition probabilities as $K_1, K_2$ and the upper-quartile transition probabilities as $H_1, H_2$. Here, we examine the performance of the detectors assuming the following reflecting boundary conditions: $P_{0,1} = 1, P_{0,i} = 0 \; \textrm{for} \; i \neq 1$ and $P_{2M,2M-1} = 1, P_{2M,i} = 0 \; \textrm{for} \; i \neq 2M$. The rest of $P$ is:
\begin{equation}
P_{ij} = \left\{ \begin{array}{ll}
K_1 & 1 \le i < M/2, j = i-1 \\
K_2 & 1 \le i < M/2, j = i+1 \\
0.5 & M/2 \le i \le 3M/2, j = i-1 \; \textrm{or} \; i+1 \\
H_1 & 3M/2 < i \le 2M-1, j = i-1 \\
H_2 & 3M/2 < i \le 2M-1, j = i+1
\end{array}\right.
\label{eqn:P_trans_mat}
\end{equation}

In the case of $M$ odd, the ranges for the indices $i,j$ would change in an obvious way. When $K_1 = H_2$ and $K_2 = H_1$, we say that the transition probabilities are symmetric, and if not, that they are asymmetric. In order to run the RT-LRT in the case of the symmetric D-T random walk, we empirically generate an average autocorrelation function of the random walk and select $p$ (and set $q=p$) so that the autocorrelation function of the symmetric D-T random telegraph matches the empirical result. From this, we also obtain the optimal $\alpha$ for the LPF of the filtered energy detector. 

\begin{figure}[!ht]
\begin{center}
\includegraphics[width=3in]{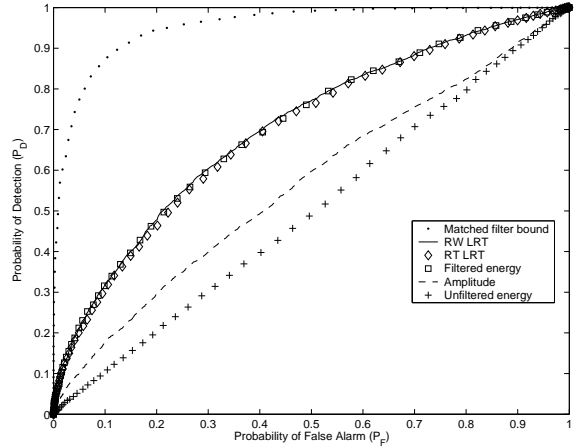}
\caption{Simulated ROC curves for Model 4 at SNR = \mbox{-39.9 dB}, $T$ = 60 s, and the symmetric random walk $K_1=K_2=H_1=H_2=0.5$ for the matched filter, D-T random walk LRT (RW-LRT), D-T random telegraph LRT (RT-LRT), filtered energy, amplitude, hybrid, and unfiltered energy detector. RW-LRT is theoretically optimal for this case.}
\label{fig:roc_mod4_sym1}
\end{center}
\end{figure}

\begin{figure}[!ht]
\begin{center}
\includegraphics[width=3in]{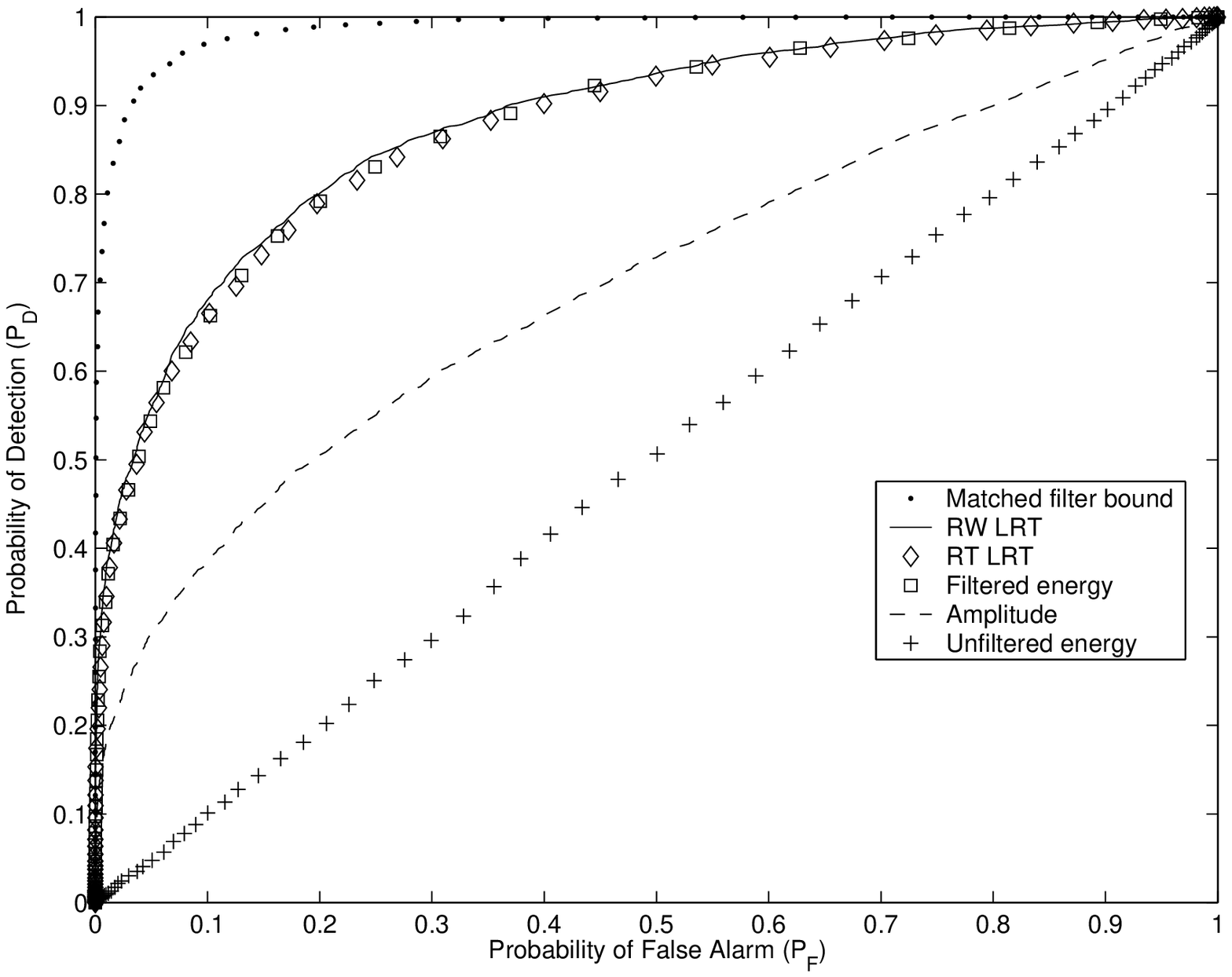}
\caption{Simulated ROC curves for Model 4 at SNR = \mbox{-37.4 dB}, $T$ = 60 s, and the symmetric random walk $K_1=H_2=0.52, K_2=H_1=0.48$. RW-LRT is theoretically optimal for this case.}
\label{fig:roc_mod4_sym2}
\end{center}
\end{figure}

\begin{figure}[!ht]
\begin{center}
\includegraphics[width=3in]{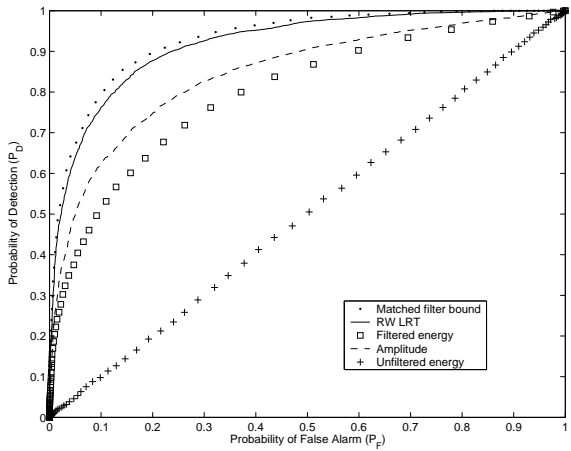}
\caption{Simulated ROC curves for Model 4 at SNR = \mbox{-41.0 dB}, $T$ = 60 s, and the asymmetric random walk $K_1=H_1=0.45$, $K_2=H_2=0.55$. RW-LRT is theoretically optimal for this case.}
\label{fig:roc_mod4_asym1}
\end{center}
\end{figure}

The ROC curves for two symmetric cases are illustrated in Figures~\ref{fig:roc_mod4_sym1} and~\ref{fig:roc_mod4_sym2}. In the former, $K_1=K_2=H_1=H_2=0.5$, while in the latter, $K_1=H_2=0.52$ and $K_2=H_1=0.48$. In both cases, the performance of the RW-LRT, RT-LRT, and filtered energy detector are all approximately the same, i.e.~the latter two detectors are nearly optimal. When the transition probabilities of the D-T random walk are asymmetric however, as in the case of Figure~\ref{fig:roc_mod4_asym1}, the D-T random walk LRT is noticeably better than the filtered energy detector.

\section{Conclusion and Discussion}
We have developed and compared optimal detectors under several single-spin MRFM signal models. While these models have to be validated through experiment, the results of this paper lend strong theoretical and practical support to the use of the simple filtered energy detector for the current MRFM single-spin research community. Indeed, we have shown that the existing baseband filtered energy detector that is in current use is approximately asymptotically optimal in the case of the symmetric D-T random telegraph model (Model 3) under the regime of low SNR, long observation times, and $p$ close to 1. The last condition can be achieved by sampling at a sufficiently fast rate as compared to the rate of random transitions. In the case of the asymmetric D-T random telegraph model, we have shown that a hybrid filtered energy/amplitude/energy detector is an approximately asymptotically optimal detector under the regime of low SNR and long observation times. We presented simulations showing that the baseband filtered energy detector has comparable performance with the optimal test statistic in the case of the symmetric D-T random walk model. We suspect that this similarity is not a coincidence, and work to verify this result analytically is ongoing. In the case of the asymmetric D-T random walk, the filtered energy detector does not perform as well as the optimal LRT. We suspect that a hybrid detector along the lines of that formulated for the D-T random telegraph will perform close to the optimal. 

\bibliographystyle{IEEEtran} 
\bibliography{IEEEabrv,mike_lib} 

\appendices

\section{Derivation of the D-T random walk LRT}
Let $y_i$ be the observations, $i=0, \ldots, N-1$. Put $y^{(i)} = (y_i, \ldots, y_0), i \ge 0$ and let $\mathcal{D} = \{0, \pm s, \ldots, \pm Ms\}$ denote the state-space. We would like to first find $f(\vec y;H_1) = f(y^{(N-1)};H_1)$, the density of the observations $\{y_i\}$ under $H_1$. From now onwards in this section, assume that the probabilities are conditioned on $H_1$ unless otherwise specified. Define $R_k(S) = P(\zeta_k=S|y^{(k-1)})$, $S \in \mathcal{D}$. Then, for $k \ge 1$,
\begin{align*}
R_k(S) &= \sum_{d \in \mathcal{D}} P(\zeta_k=S|\zeta_{k-1}=d)
	P(\zeta_{k-1}=d|y^{(k-1)}) \\
	&= \frac{ \sum_{d \in \mathcal{D}} P(\zeta_k=S|\zeta_{k-1}=d)
		f_w( y_{k-1} - d ) R_{k-1}(d) }{
			\sum_{i \in \mathcal{D}} f_w( y_{k-1} - i ) R_{k-1}(i) } 
\end{align*}
which results in
\begin{equation}
\overrightarrow{R}_k =
	\left( \overrightarrow{W}_{k-1} \cdot \overrightarrow{R}_{k-1} 
		\right)^{-1} Q \left(\overrightarrow{W}_{k-1} * 
			\overrightarrow{R}_{k-1} \right)
\label{eqn:rw_Rk}
\end{equation}
for $k \ge 1$, $\overrightarrow{R}_0 = \frac{1}{2}(e_M + e_{M+2})$ since $\zeta_0$ is equally likely to be either $\pm s$. Next, for $k \ge 1$,
\begin{align*}
f( y_k | y^{(k-1)} ) &= \sum_{d \in \mathcal{D}} P(\zeta_k=d|y^{(k-1)}) 
	f_w( y_k - d ) \\
	&= \sum_{d \in \mathcal{D}} R_k(d) f_w( y_k - d ) \nonumber \\
	&= \overrightarrow{R}_k \cdot \overrightarrow{W}_k 
\end{align*}
and $f(y_0)=\overrightarrow{R}_0 \cdot \overrightarrow{W}_0$. This leads to
\begin{align}
f( y^{(N-1)} ) &= f( y_{N-1} | y^{(N-2)} )
	\cdots f( y_1 | y_0 ) f( y_0 ) \nonumber \\
	&= \prod_{k=0}^{N-1} \overrightarrow{R}_k \cdot \overrightarrow{W}_k
\label{eqn:rw_fH1}
\end{align}

The density of the observations $\{y_i\}$ under $H_0$ is
\begin{align}
f( y^{(N-1)}; H_0 ) &= \prod_{k=0}^{N-1} \mathcal{N}(y_k;0, \sigma^2) \nonumber\\
	&= \prod_{k=0}^{N-1} f_w( y_k ) \nonumber \\
	&= \prod_{k=0}^{N-1} \left( \vec{e}_{M+1} \cdot \overrightarrow{W}_k \right)
\label{eqn:rw_fH0}
\end{align}
and the LRT is then, using (\ref{eqn:rw_fH1}) and (\ref{eqn:rw_fH0}):
\begin{align}
\Lambda(\vec y) &= \frac{ f( \vec y; H_1 ) }{ f( \vec y; H_0 ) } \nonumber \\
	&= \prod_{k=0}^{N-1} \frac{
		\overrightarrow{R}_k \cdot \overrightarrow{W}_k }{
			\vec{e}_{M+1} \cdot \overrightarrow{W}_k }
\label{eqn:rw_lrt}
\end{align}

\section{Approximate second-order expansion of the D-T random telegraph LRT and comparison to the filtered energy test statistic} 
Let $\mathcal{T}_1(\vec y)$ denote log LRT of the D-T random telegraph in (\ref{eqn:2state_lrt}), and $\mathcal{T}_2(\vec y)$ the filtered energy detector in (\ref{eqn:det_fe}). Let us analyze the two test statistics under the regime of low SNR ($\left|\frac{A}{\sigma}\right| \ll 1$) and long observation times ($N \gg 1$). We want to obtain the approximate second-order expansion of $\mathcal{T}_1(\vec y)$. Write $\mathcal{T}_1(\vec y) \simeq L_1 + L_{2a} + L_{2b} + \textrm{h.o.t.}$, where $L_1$ are the 1st order terms, $L_{2a}$ are the 2nd order terms consisting of $y_j y_k$ where $j < k$, $L_{2b}$ are the 2nd order terms of the form $y_k^2$, and ``h.o.t.'' are the higher order terms. Define: $T_k(S) = R_k(S)e^{\frac{S}{\sigma^2} y_k}, \; S \in \{\pm A\}$ and $\theta_k = \frac{T_k(A)}{T_k(A) + T_k(-A)}$, for $k \ge 0$. From (\ref{eqn:R_simpler}), a recursive equation for $\theta_k$ can be derived. Its approximate solution is
\begin{align}
\theta_k &= \beta_k + \frac{qA}{\sigma^2} \sum_{j=0}^k \xi_{kj} y_j,
k \ge 0 \; \textrm{where} \nonumber \\
\beta_k &= \frac{1-q}{1-r} + \left( \frac{1}{2} - \frac{1-q}{1-r} \right) r^k,
	\; k \ge 0 \nonumber \\
\xi_{kj} &= \frac{ 2(1-q)r^{k-j} + (2q-r-1)r^k }{ 1-r }, \; 0 \le j \le k-1
	\nonumber \\
\xi_{kk} &= \frac{2(1-q)}{1-r} + \frac{r^k(2q-r-1)}{1-r} = 2\beta_k, \;
	k \ge 0
\label{eqn:thetak_soln}
\end{align} 
and $r = p + q - 1$. Note that $p,q \in (0,1) \Rightarrow |r| < 1$. Define $s_k = \frac{A}{\sigma^2}y_k$. Then,
\begin{align}
\mathcal{T}_1 \simeq &
	\sum_k \bigg\{ \big[s_k(2R_k(A)-1)+\frac{1}{2}s_k^2\big] \, - 
	\nonumber \\
		&\quad\quad \frac{1}{2}\big[s_k(2R_k(A)-1)+\frac{1}{2}s_k^2\big]^2
	\bigg\}
\label{eqn:T1_raw}
\end{align}
By solving for $R_k(A)$ in terms of $\theta_k$ and using (\ref{eqn:thetak_soln}) in (\ref{eqn:T1_raw}), one can sort out the terms and obtain expressions for $L_1$, $L_{2a}$ and $L_{2b}$. Let $C_m = \frac{p-q}{2-p-q}$. This number gives an indication of the mismatch in $p$ and $q$. 
\begin{align}
L_1 &= 
	\frac{A}{\sigma^2} C_m \sum_k (1-r^k) y_k 
	\label{eqn:coeff_yk} \\
L_{2a} &= 2q \left(\frac{A}{\sigma^2}\right)^2 \sum_k \sum_{j=0}^{k-1} \left[
	\frac{2(1-q)}{1-r} r^{k-j} - r^k C_m  \right] y_j y_k
	\label{eqn:coeff_yjyk}  \\
L_{2b} &= \left(\frac{A}{\sigma^2}\right)^2 \sum_k \bigg\{
	4r\left(\frac{1-q}{1-r}\right)^2 + 
	2\frac{(q-r)(1-q)}{(1-r)^2} \nonumber \\
	& \quad\quad - C_m(2q+C_m)r^k + \frac{1}{2}C_m^2 r^{2k} \bigg\} y_k^2
	\label{eqn:coeff_yk2}
\end{align} 

When $p=q$, $C_m=0$. This simplifies $\mathcal{T}_1(\vec y)$ considerably. From (\ref{eqn:coeff_yk})-(\ref{eqn:coeff_yk2}),
\begin{align}
\mathcal{T}_{1s}(\vec y) 
	&= 2p \left(\frac{A}{\sigma^2}\right)^2 \bigg\{
		\sum_{k=1}^{N-1} \sum_{j=0}^{k-1} (2p-1)^{k-j} y_j y_k \nonumber\\
	& + \sum_{k=0}^{N-1} \left(1-\frac{1}{4p}\right) y_k^2 \bigg\}
\label{eqn:lrt_symm}
\end{align}
where $\mathcal{T}_{1s}$ denotes $\mathcal{T}_1$ when the transition probabilities are symmetric.

Next, let us obtain an expression for $\mathcal{T}_2(\vec y)$. For sufficiently large $N$, it can be shown that 
\begin{equation}
\mathcal{T}_2(\vec y) \simeq D \left\{
	\sum_{k=1}^{N-1} \sum_{j=0}^{n-1} \alpha^{k-j} y_j y_k +
	\frac{\alpha}{1+\alpha} \sum_{k=0}^{N-1} y_k^2 \right\}
\label{eqn:filt_energy_ts}
\end{equation}
where $D=\frac{1-\alpha^2}{2\alpha}$ is a constant. Note that $D$ plays no role in the performance of the test statistic. Comparing (\ref{eqn:lrt_symm}) and (\ref{eqn:filt_energy_ts}), we see that they are nearly identical in form if $\alpha = 2p-1$. The summation of the cross-terms will be the same, but the coefficient of the energy term will be $(1-\frac{1}{4p})$ in the case of $\mathcal{T}_{1s}$ and $(1-\frac{1}{2p})$ in the case of $\mathcal{T}_2$. However, the contribution of $-\frac{1}{4p}\sum_k y_k^2$ to $\mathcal{T}_2$ is not as significant as the summation of the cross-terms. Now, $E[\sum_{k=0}^{N-1}y_k^2;H_1]-E[\sum_{k=0}^{N-1}y_k^2;H_0]=NA^2$, and it can be shown that for large $N$,
\begin{align}
& E\bigg[\sum_{k=1}^{N-1} \sum_{j=0}^{k-1} \alpha^{k-j} y_j y_k;H_1\bigg] -
E\bigg[\sum_{k=1}^{N-1} \sum_{j=0}^{k-1} \alpha^{k-j} y_j y_k;H_0\bigg]
\nonumber \\
& \simeq G A^2(N-1)
\label{eqn:crossterms_diff}
\end{align}
where $G = \frac{\alpha(2p-1)}{1-\alpha(2p-1)}$. When $\alpha=2p-1$, $G = \frac{(2p-1)^2}{1-(2p-1)^2} = \frac{1}{4(1-p)} + \frac{1}{4p}-1$. For $p$ close to 1, $G \gg 1$, and $G A^2(N-1) \gg A^2 N$. So to the first moment, the additional $-\frac{1}{4p}\sum_k y_k^2$ to $\mathcal{T}_2$ in order to make it equal to $\mathcal{T}_{1s}$ does not represent a significant difference. When $p \approx 1$, we expect that the performance of the filtered energy detector and the LRT to be similar.

It is possible to obtain an approximation to the second-order expansion of the D-T random telegraph LRT by combining the filtered energy, amplitude, and energy statistics. Firstly, for large $N$, the LRT is approximately
\begin{align}
\widetilde{\mathcal{T}}_1(y) &= C \bigg\{ 
	\underbrace{\frac{(p-q)\sigma^2}{4q(1-r)A}}_{C_{\textrm{I}}} \sum_k y_k +
	\sum_k \sum_{j<k} r^{k-j} y_j y_k \nonumber \\
	&\quad + \bigg[\frac{1}{2} + 
		\underbrace{\frac{r(1-q)}{2q(1-r)}}_{C_{\textrm{II}}} \bigg] 
		\sum_k y_k^2 
	\bigg\}
\label{eqn:Ty_approx2}
\end{align}
where $C = 4q\frac{1-q}{1-r}\left(\frac{A}{\sigma^2}\right)^2$ is constant. Let $\mathcal{T}_h(y)$ denote the hybrid detector that is composed of the linear combination of the amplitude, filtered energy, and unfiltered energy statistics. We see that in the filtered energy statistic (\ref{eqn:filt_energy_ts}), the ratio of the energy terms to the cross terms is $\frac{\alpha}{1+\alpha}$. For $\alpha \approx 1$, this is roughly 1/2. The idea is to add the energy and amplitude statistics so that all three statistics are in the same ratio as in (\ref{eqn:Ty_approx2}). So, put:
\begin{align}
\mathcal{T}_h(y) &= \mathcal{T}_2(y) + 
	\frac{1-\alpha^2}{2\alpha}\left[ C_{\textrm{I}} \sum_k y_k
		+ C_{\textrm{II}} \sum_k y_k^2 \right] \nonumber \\
	&= \mathcal{T}_2(y) + \frac{1-\alpha^2}{2\alpha} C_{\textrm{I}} 
		\sum_k y_k +
		\frac{1-\alpha^2}{2\alpha} C_{\textrm{II}} \sum_k y_k^2
\label{eqn:hybrid_ts}
\end{align}

We expect the approximation $\widetilde{\mathcal{T}}_1(y)$ to have performance that is similar to the generalized LRT when the number of samples $N$ is large. Since $\mathcal{T}_h(y)$ is equivalent to $\widetilde{\mathcal{T}}_1(y)$, the same follows for the hybrid detector.

\end{document}